\documentclass[lettersize,journal]{IEEEtran}
\usepackage{amsmath,amsfonts}
\usepackage{algorithmic}
\usepackage{algorithm}
\usepackage{array}
\usepackage[caption=false,font=normalsize,labelfont=sf,textfont=sf]{subfig}

\usepackage{textcomp}
\usepackage{stfloats}
\usepackage{url}
\usepackage{verbatim}
\usepackage{hyperref}
\usepackage{graphicx}
\usepackage{cite}
\usepackage{color}
\usepackage{amsfonts,amsmath,amssymb}
\usepackage{amsmath}
\usepackage{algorithm} 
\usepackage{amsmath, amssymb}
\usepackage{multirow}
\usepackage{tabularx}
\usepackage{booktabs} 
\usepackage{float} 
\usepackage{pdflscape} 

\begin{document}

\title{Multimodal Trustworthy Semantic Communication for Audio-Visual Event Localization}

\author{Yuandi Li,~\IEEEmembership{Memember,~IEEE,} Zhe Xiang, Fei Yu,~\IEEEmembership{Memember,~IEEE,} Zhuoran Zhang,\\ Zhangshuang Guan, Hui Ji, Zhiguo Wan,  Cheng Feng

\thanks{The corresponding authors: Fei Yu. Yuandi Li, Zhe Xiang and Hui Ji are with Jiangsu University, Zhenjiang, 212013, China; Fei Yu, Zhuoran Zhang and Zhiguo Wan are with Zhejiang Lab, Hangzhou, 311121, China; Zhuangshuo Guan is with the College of Computer Science and Technology, Zhejiang University, Hangzhou, 310027, China; Cheng Feng is with School of Computer and Control Engineering, Northeast Forestry University, Harbin, 150040, China.
 }}
\markboth{Journal of \LaTeX\ Class Files,~Vol.~14, No.~8, August~2021}%
{Shell \MakeLowercase{\textit{et al.}}: A Sample Article Using IEEEtran.cls for IEEE Journals}


\maketitle
\begin{abstract}
The exponential growth in wireless data traffic, driven by the proliferation of mobile devices and smart applications, poses significant challenges for modern communication systems. Ensuring the secure and reliable transmission of multimodal semantic information is increasingly critical, particularly for tasks like Audio-Visual Event (AVE) localization. This letter introduces MMTrustSC, a novel framework designed to address these challenges by enhancing the security and reliability of multimodal communication.

MMTrustSC incorporates advanced semantic encoding techniques to safeguard data integrity and privacy. It features a two-level  coding scheme that combines error-correcting codes with conventional encoders to improve the accuracy and reliability of multimodal data transmission. Additionally, MMTrustSC employs hybrid encryption, integrating both asymmetric and symmetric encryption methods, to secure semantic information and ensure its confidentiality and integrity across potentially hostile networks. Simulation results validate MMTrustSC's effectiveness, demonstrating substantial improvements in data transmission accuracy and reliability for AVE localization tasks. This framework represents a significant advancement in managing intermodal information complementarity and mitigating physical noise, thus enhancing overall system performance.

\end{abstract}

\section{Introduction}
\IEEEPARstart{T}{he} rapid growth of wireless data traffic, driven by the proliferation of mobile devices and the increasing demand for smart services, presents substantial challenges in modern communication systems. Traditional communication paradigms focus on converting data into bits for transmission and ensuring precise bit recovery at the receiver. This approach relies heavily on high channel quality and signal-to-noise ratios to maintain data integrity. However, semantic communication has emerged as a promising alternative, aiming to directly transmit and retrieve the meaning of the content, thereby enhancing robustness under complex and noisy channel conditions~\cite{qin2019deep}.

Most existing research in semantic communication has concentrated on single-modal tasks~\cite{xie2021task,xie2022task}. However, in multimodal tasks, where multiple types of data such as audio and video are integrated, ensuring the complementarity of inter-modal information and enhancing semantic richness become crucial. The transmission of multimodal data is particularly susceptible to physical noise, which exacerbates inter-modal interference and negatively impacts task performance and system effectiveness.

Traditional methods of semantic communication often depend on joint source-channel coding, which, while effective, introduces potential risks, such as information leakage and security vulnerabilities~\cite{xie2024hybrid,gu2023semantic}. These risks are especially pronounced when handling concurrent audio and video data, as the approach can struggle to maintain data integrity and privacy across different modalities. Additionally, the process of encoding and decoding in semantic communication introduces two types of noise: semantic noise from information loss during the encoding process and channel noise from physical disturbances during transmission. These factors contribute to semantic loss and intermodal interference, undermining the reliability and effectiveness of multimodal tasks.

Encryption algorithms are crucial to ensure that information transmitted over public channels remains secure, even if intercepted, thus safeguarding data confidentiality. For nearly all standard encryption algorithms, successful decryption depends not only on possessing the correct decryption key but also on the integrity of the ciphertext. However, in traditional semantic communication, where information is transmitted in the real number domain, minor errors during transmission can still yield relevant semantic features. This challenges for applying standard encryption algorithms to effectively protect semantic information, as the decoder's output may differ from the original information.

To address these challenges, we propose \textbf{M}ultimodal \textbf{T}rustworthy \textbf{S}emantic \textbf{C}ommunication (\textbf{MMTrustSC}): a framework specifically designed for the multimodal task--Audio-Video Event (AVE) localization. MMTrustSC integrates advanced semantic encoding techniques to ensure data integrity and privacy throughout transmission. It features a two-level coding scheme that combines conventional channel encoders with an outer layer of error-correcting codes, improving reliability by allowing minor errors to be corrected to ensure accurate outputs.

Moreover, MMTrustSC incorporates a hybrid encryption mechanism to protect semantic information. Initially, a secure session key is established between the sender and the receiver using public-key cryptography. This session key is then employed for symmetric encryption of the transmitted information, combining the benefits of both cryptographic methods—efficient key exchange from public-key cryptography and high-performance encryption from symmetric-key cryptography.

The simulation results demonstrate that MMTrustSC effectively preserves the fidelity of the multimodal semantic information transmitted, enabling accurate AVE localization by seamlessly integrating and processing audio and visual data. By embedding encryption protection and semantic error correction mechanisms, MMTrustSC addresses security and noise challenges, significantly enhancing system robustness, reliability, and ensuring the integrity and privacy of multimodal data.

\textbf{Summary of Novel Contributions:}
\begin{itemize}
    \item \textbf{MMTrustSC Framework:} Introduces a novel framework for secure and reliable multimodal communication, specifically targeting AVE. The framework enhances robustness through advanced semantic encoding techniques that ensure data integrity and privacy.
    \item \textbf{Two-level coding scheme:} Implements a two-level coding approach that integrates error-correcting codes with conventional encoders, improving the reliability and accuracy of multimodal data transmission.
    \item \textbf{Hybrid Encryption:} Utilizes a hybrid encryption strategy that combines asymmetric and symmetric encryption to secure semantic information, ensuring confidentiality and integrity even in potentially hostile network environments.
\end{itemize}

\section{Methodology}
\begin{figure*}
    \centering
    \includegraphics[width=0.85\linewidth]{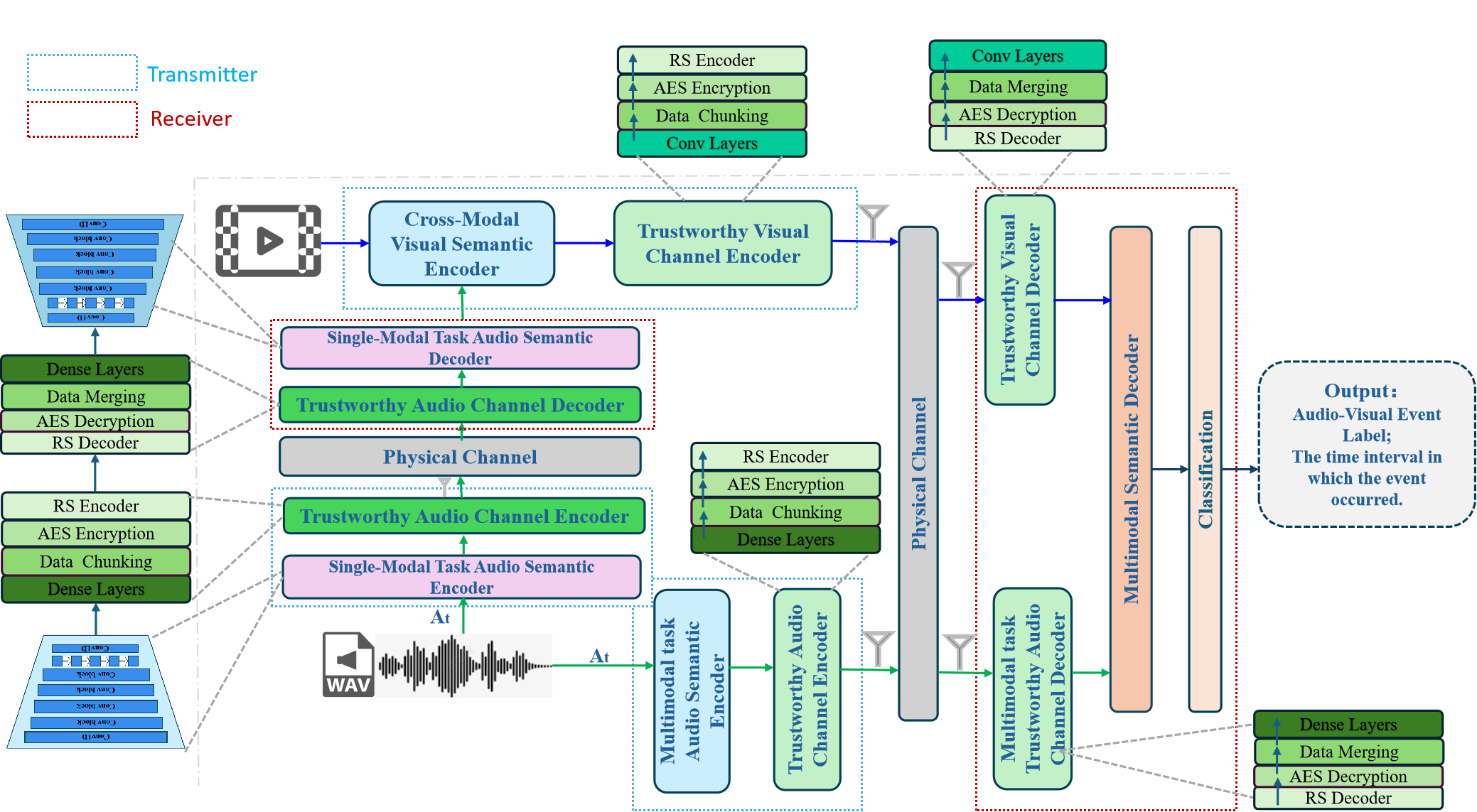}
    \caption{The overall framework of the mulimodal trustworthy semantic communication for the multimodal task.}
    \label{fig:framework}
\end{figure*}
In this letter, we focus on the design of a semantic communication system for multimodal tasks in a multi-user setting. Specifically, we address the semantic communication of multimodal data streams captured by multimodal sensors in practical application scenarios. In our design, there are two transmitting users: one user with audio data and a single antenna, and another user with video data and two antenna. The multimodal receiving end is equipped with $M$ antennas. It is straightforward to expand the network to handle inputs from multiple video and audio sources. Furthermore, we design a deep neural network (DNN) for the semantic communication system to serve the task, where all models can be trained in the cloud and then broadcast to users. As shown in Fig.~\ref{fig:framework}, our disigned multimodal trustworthy semantic communication framework contains semantic encoder modules, channel encoder modules, channel decoder modules, and semantic decoder modules.

\subsection{Transmitter}
In MMTrustSC's transmitter, there are several components: audio semantic encoders (for single-modal task semantic encoding and multimodal task semantic encoding), visual semantic encoders, and trustworthy visual (audio) channel encoders.

\subsubsection{Audio Semantic Encoder}
This paper focuses on multimodal Audio-Visual Event localization (AVE). Two users transmit audio and video data through semantic and channel encoding over a physical channel to a multi-antenna receiver. The receiver performs decoding to identify event labels and time intervals, enhancing and complementing multimodal data. The audio user also sends audio information to the video user, using single-modal and multimodal task audio semantic encoder-decoder modules. 
\noindent \textbf{For single-modal task audio semantic encoder:} The goal is to enhance the semantic features of video using audio signals transmitted from an audio-equipped user to a video-equipped user through a physical channel. The audio semantic encoder-decoder focuses on accurately recovering the audio signal. The encoder model is denoted by:
\begin{equation}
    \mathbf{A}_{single}=\mathcal{SE}_{a,1}(\mathcal{S}^a;\mathbf{\Theta}_{a,1}),
\end{equation}
where $\mathbf{\Theta}_{a,1}$ represents the trainable parameters, which consists of a 1D convolution with \( C \) channels and a kernel size of 7, followed by \( B \) convolution blocks. Each block includes a residual unit and a down-sampling layer with a strided convolution, where the kernel size \( K \) is twice the stride \( S \). The residual unit has two convolutions with a kernel size of 3 and a skip-connection. Channels double with each down-sampling. The convolution blocks are followed by a two-layer LSTM for sequence modeling and a final 1D convolution layer with a kernel size of 7 and \( D \) output channels. Following previous work, we use \( C = 32 \), \( B = 4 \), and strides of (2, 4, 5, 8). ELU is used as the activation function, with either layer normalization or weight normalization.

\noindent \textbf{For multimodal task audio semantic encoder:} 
In multimodal tasks, the goal is to transmit audio features through a physical channel to support the completion of the multimodal AVE tasks. Given the audio segment sequence \(\{S_t^a\}_{t=1}^T\), the encoder model is denoted by:
\begin{equation}
    \mathbf{A}_{multi}=\mathcal{SE}_{a,2}(\mathcal{S}^a;\mathbf{\Theta}_{a,2}),
\end{equation}
where $\mathbf{\Theta}_{a,2}$ represents the trainable parameters. We first use the pre-trained VGG-19 model~\cite{VGG19} to extract audio features \(\mathbf{A}_{multi} = \{\mathbf{a}_t\}_{t=1}^T \in \mathbb{R}^{T \times d_a}\), where \(d_a\) is the feature dimension of the audio segments. Next, an LSTM network with 512 units and \textit{tanh} activation performs semantic encoding of these audio features. This encoded semantic representation is then used as input for the channel encoding stage.
\subsubsection{Cross-Modal Visual Semantic Encoder}
At the video-equipped user end, the received audio signal \(\mathcal{S}^{a*}\) is accurately recovered during the single-modal task decoding stage. This recovered signal is then used in the cross-modal semantic encoder to enhance visual features.

The cross-modal visual semantic encoder is defined as:
\begin{equation}
   \mathbf{V} = \mathcal{SE}_v(\mathcal{S}^v, \mathcal{S}^{a*}; \mathbf{\Theta}_v), 
\end{equation}
where \(\mathbf{\Theta}_v\) are the trainable parameters of the encoder. Specifically, visual features \(\mathcal{V} = \{\mathbf{v}_t\}_{t=1}^T\) and the features of the received audio signal \(\hat{\mathbf{A}}_{single} = \{\mathbf{a}_t^{*}\}_{t=1}^T\) are extracted from the video using a pre-trained VGG-19 model. The AGVA method~\cite{ave} is applied to compute attention-weighted visual features \(\mathcal{V}^{att} = \{\mathbf{v}_t^{att}\}_{t=1}^T\), which emphasize regions relevant to the audio. Attention weights \(\alpha_t\) are calculated as:
$\alpha_t = \sigma\left(W_f \tanh(W_v^1 M_v (\mathbf{v}_t) + W_a^1 M_a (\mathbf{a}^{*}_t))\right)$, where \(M_v\) and \(M_a\) project visual and audio features into a shared space, and \(W_v^1\), \(W_a^1\), and \(W_f\) are trainable parameters. This process ensures that the visual features are aligned with the audio information.

\subsubsection{Trustworthy Audio Channel Encoder}
\label{sec:trustChannel}


In the context of single-modal task semantic transmission for audio, channel encoding primarily comprises four parts: two blocks in the data preprocessing stage (dense layer and data chunking) and two blocks in the trustworthy encoder (AES Encryption~\cite{aes} and RS Encoder~\cite{rs}).

\noindent \textbf{Trustworthy Encoder:} 
To construct the trustworthy encoder, we introduces a double-layer coding architecture. By appending an outer layer of error-correcting code over the data chunking layer, it enables a secondary correction for potential low-weight errors arising from conventional channel decoder. This design ensures the accuracy of the output decoded information, leading to a robust channel encoder. As a result, it solidifies the essential groundwork for integrating cryptographic algorithms within the framework of semantic communication.

In this paper, we use hybrid encryption to facilitate the confidential transmission of semantic information. In more detail, we first establish a shared secret session key between the sender and receiver using public-key cryptography, followed by symmetric encryption of the transmitted information with this secret session key. Suppose the semantic decoder holds the key pairs $(pk,sk)$ of the Elliptic Curve Cryptography (ECC) algorithm~\cite{ecc}, then the complete algorithm works as Algorithm 1. The output of trustworthy encoder is denoted:
\begin{equation} \mathbf{X}_a=\mathcal{CE}_a(\mathbf{A};\mathbf{\Phi}_a),
\end{equation}
where $\mathbf{A}\in \{\mathbf{A}_{single},\mathbf{A}_{multi}\}$; $\mathbf{\Phi}_a$ is the trainable parameters.
\begin{algorithm}
\caption{Audio (Visual) Sender Ciphertext Generation}
\label{alg:audio_sender}
\begin{algorithmic}[1]
\REQUIRE Audio message $m$, 16-bit use key $usekey$, ECC public key $pk$.
\ENSURE Ciphertext $c$.
\STATE $k \leftarrow \text{aes\_set\_encrypt\_key}(userkey_v)$;
\STATE $c1 \leftarrow \text{ecc\_enc}(pk, k)$;
\STATE $m \leftarrow \text{process With Data Chunking Model}(m)$;
\STATE $c2 \leftarrow \text{aes\_cbc\_enc}(k, m)$;
\STATE $h \leftarrow \text{sha3}(m)$;
\STATE $c \leftarrow (c1, c2, h)$;
\RETURN $c$
\end{algorithmic}
\end{algorithm}

Notice that, sharing the AES session key within this algorithm operates completely independently of the semantic communication itself. This means that through pre-arrangement or other agreed-upon methods, participants can share a secure secret value before the onset of semantic communication. This shared secret will subsequently be utilized for the generation of the AES session key. We then apply Reed-Solomon (RS) encoding to each of the data streams, specifying the number of roots in the generator polynomial (nroots) as 1280 and using an extension field of GF($2^{18}$). Convert the aforementioned processed real feature numbers into complex numbers.


\subsubsection{Trustworthy Visual Channel Encoder} 
\noindent Similar to the \textbf{Trustworthy Encoder} described in Section~\ref{sec:trustChannel} for the trustworthy audio channel encoder, the trustworthy visual channel encoder also ensures reliable encoding and decoding of information. However, before encoding the trustworthy channel, the semantically encoded features undergo pre-processing. Specifically, features are processed through two Conv2D layers to obtain $\mathbf{V}^{'}$. The primary purpose of this preprocessing step is to further refine the features through cross-modal visual semantic analysis, enhancing their representation and alignment with the audio features. Thus, the trustworthy visual channel encoder can be represented as follows.
\begin{equation}
    \mathbf{X}_v=\mathcal{CE}_v(\mathbf{V}^{'};\mathbf{\Phi}_v),
\end{equation} 
where $\mathbf{\Phi}_v$ represents the trainable parameters.

\subsection{Receiver}

\subsubsection{Trustworthy Audio (Visual) Channel Decoder}
At the receiver end for multi-modal tasks, signals are transmitted through a physical channel and then processed via signal detection. 

\noindent \textbf{Signal Detection:}
We adopt the Zero-Forcing Linear Minimum Mean Square Error (ZF-LMMSE) detector, which combines the advantages of Zero-Forcing (ZF) and Linear Minimum Mean Square Error (LMMSE). Its goal is to eliminate multipath interference while reducing noise amplification.We define \( \mathbf{Y}\) as the received signal vector, \( \mathbf{X}\) as the transmitted signal vector, \(\mathbf{H}\) as the channel matrix, and \(\mathbf{N}\) as the additive white Gaussian noise vector. The received signal can be expressed as:
\begin{equation}
\mathbf{Y} = \mathbf{HX} + \mathbf{N}.
\end{equation}
In the ZF-LMMSE detector, our aim is to find a detection matrix \(\mathbf{W}\) such that the output approaches the true transmitted signal \(\hat{\mathbf{X}}\).
The detection matrix \(\mathbf{W}\) for the ZF-LMMSE detector can be expressed as:
\begin{equation}
\hat{\mathbf{X}} = \mathbf{WY} = \left( \mathbf{H}^H\mathbf{H} + \frac{\sigma_n^2}{\sigma_x^2}\mathbf{I} \right)^{-1}\mathbf{H}^H\mathbf{Y},
\end{equation}
where \(\mathbf{H}^H\) is the conjugate transpose of the channel matrix \(\mathbf{H}\); \(\sigma_n^2\) is the variance of the noise; \(\sigma_x^2\) is the variance of the transmitted signal; \(\mathbf{I}\)  is the identity matrix.

\noindent \textbf{Trustworthy Decoder:}
After signal detection, the estimated complex signals are first converted into real signals and then undergo RS decoding through the trustworthy decoder. Following this, AES decryption is performed, and the data is then converted into audio and visual feature information by data merging.

After receiving the message and performing the RS decoding, the semantic decoder retrieves the encrypted message \( c \). The message is then recovered as follows:
\begin{algorithm}
\caption{Audio (Visual) Message Recovery Process}
\label{alg:message_recovery}
\begin{algorithmic}[1]
\REQUIRE Encrypted message $c$, ECC private key $sk$.
\ENSURE Decrypted message $m$.
\STATE $c1 \leftarrow \text{First 160 bits of } c$;
\STATE $h \leftarrow \text{Last 128 bits of } c$;
\STATE $c2 \leftarrow \text{Remaining middle bits of } c$;
\STATE $k' \leftarrow \text{ecc\_dec}(sk, c1)$;
\STATE $m' \leftarrow \text{aes\_cbc\_dec}(k', c2)$; 
\IF {$\text{sha3}(m') = h$}
    \STATE \textbf{Decryption successful} 
\ELSE
    \STATE \textbf{Request retransmission}
\ENDIF
\RETURN $m$
\end{algorithmic}
\end{algorithm}

\subsubsection{Multimodal Semantic Decoder}
With the combined audio and visual semantic information, we utilize the Positive Sample Propagation Network~\cite{psp} as the semantic decoder. This network is responsible for merging the audio and visual semantic data and accurately localizing the multimodal event localization.
\begin{equation}
    \textit{Multimodal\_Task}=\mathcal{SD}
    (\hat{\mathbf{A}}_{multi},\hat{\mathbf{V}};\mathbf{\phi}),
\end{equation}
where $\phi$ is the trainable parameters. 

\subsubsection{Loss Function}
In the AVE task, we utilize a combined objective function to optimize network performance. First, we utilize the cross-entropy loss \(L_{CE}\) to measure the difference between the true category \(a\) and the predicted category \(\hat{a}\). The formula is as follows.
\begin{equation}
    L_{CE}(p_{av}, \hat{p}_{av};\mathbf{\Theta}_{a,1},\mathbf{\Theta}_{a,2}, \mathbf{\Theta}_{v}, \mathbf{\Phi}_{a},\mathbf{\Phi}_{v},\phi) = -p_{av} \log(\hat{p}_{av}),
\end{equation}
where \(p_{av}\) is the probability of the true audio-visual event and \(\hat{p}_{av}\) is the probability of the predicted event category. Minimizing this loss improves the accuracy of event classification.

Furthermore, we incorporate the loss of similarity between audio-visual pairs \(L_{AVPS}\)~\cite{psp}. This loss improves the consistency of the audio and visual features by calculating the mean squared error between the similarity vector \(S\) of audio and video features and the label vector \(G\). 

The final objective function combines these two components:
\begin{equation} 
L_{fully} = L_{CE} + \lambda L_{AVPS}, 
\end{equation}
where \( \lambda \) is a hyperparameter that balances classification accuracy and consistency of features. This combined loss function ensures accurate event classification while optimizing the alignment of audio and visual features, thereby improving overall localization performance.

\section{Experiments}

\subsection{Experiment Settings}
\subsubsection{Dataset} 
In this work, we use the AVE subset of the Audioset dataset~\cite{tian2018audio}, which includes 4,143 videos across 28 event types with temporal annotations. Each video contains at least one audio-visual event lasting a minimum of 2 seconds, covering various domains such as human activities, musical performances, and vehicle sounds. Our primary task is to localize events by identifying the event type for each one-second segment in synchronized video and audio streams.

\subsubsection{Evaluation Metrics}
In this work, we focus on the multimodal task of audiovisual event localization, using accuracy as the primary evaluation metric. We predict the category label for each segment in a fully supervised setting and assess performance across different Signal-to-Noise Ratios (SNR) and channel states.
\begin{figure*}[t]
    \centering
    \includegraphics[width=\textwidth]{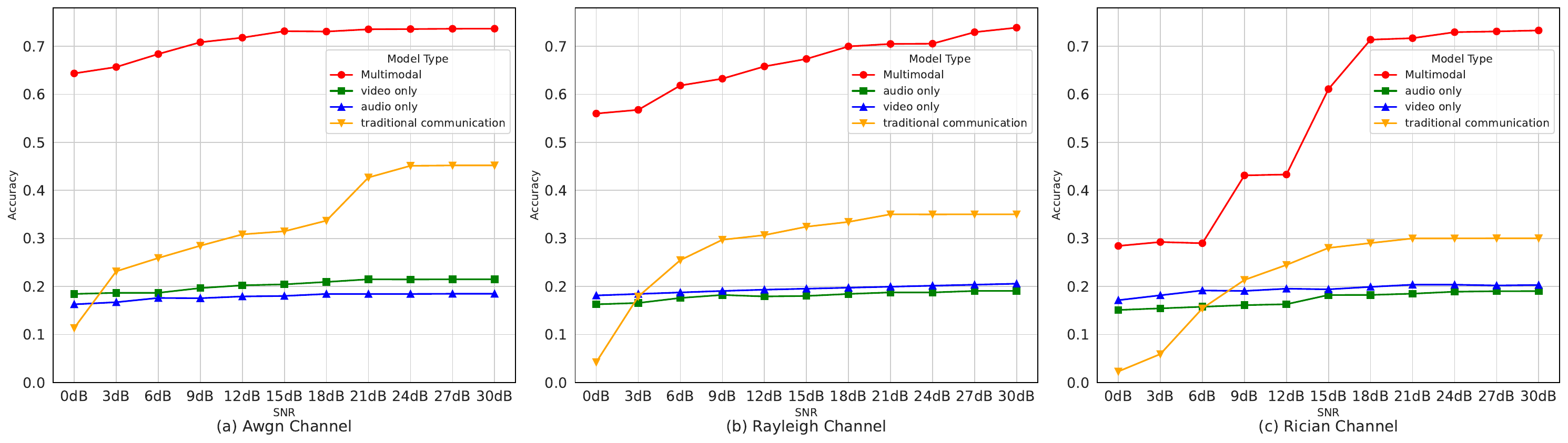} 
    \caption{The accuracy for various testing channels based on trained models.}
    \label{fig:accuracy}
\end{figure*}
\begin{figure*}[!h]
    \centering
    \includegraphics[width=0.85\textwidth]{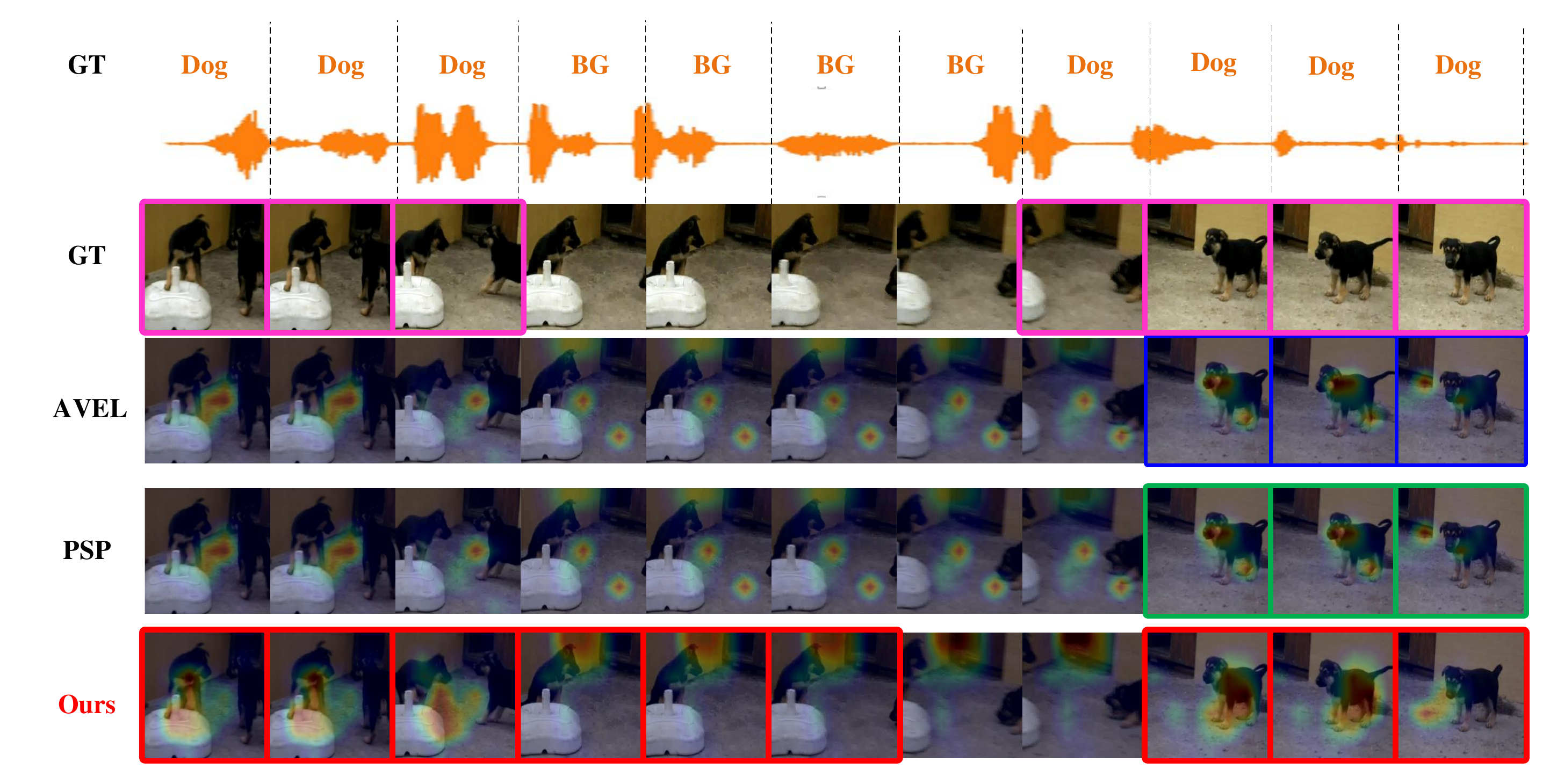} 
    \caption{Audio-Visual Event Localization Heatmap. The illustrated video segment features the audio wave representing the sound waveform, with BG denoting the background. Only the first three and the last four segments contain visual and audio signals pertaining to the audio-visual event (dog barking). The diagram showcases predictions made by three methods: PSP, AVEL, and our proposed approach.}
    \label{fig:heatmap}
\end{figure*}
\subsubsection{Detail Implementations}
To ensure a fair comparison with the original AVE study~\cite{ave}, we utilize the same visual and audio features. Visual features are extracted using VGG-19 pre-trained on ImageNet, and audio features are obtained by converting raw audio into log-mel spectrograms, processed through a VGG-like network pre-trained on AudioSet. Audio features are then passed through a bidirectional LSTM with a hidden size of 128 per direction, while video features undergo cross-modal attention and are processed similarly with an input size of 512 dimensions. Post-LSTM, video features are reduced to 256 dimensions via 2D convolution, and audio features are upscaled to 256 dimensions. These features are segmented, encoded using AES, encrypted with RS encoding, transmitted, decoded, and decrypted at the receiver end, followed by deep integration for high-level abstraction in classification tasks. The video processing path involves two 2D convolutional layers with ELU activation, scaling from 256 to 512 dimensions, while the audio path includes three linear layers with ReLU activation, progressively increasing to 512 dimensions. The model is trained for 300 epochs with the ScheduledOptim optimizer, an initial learning rate of $3e\text{-}4$ (decreasing by 0.1 every 10 epochs), and a batch size of 64. Key hyperparameters, $\lambda = 100$ and $\tau_1 = 0.099$, were determined through extensive experimentation.
All experiments were conducted on a single NVIDIA RTX 2080Ti GPU\footnote{The code for our method is available at \url{https://github.com/dimlight13/MU_SC_for_VQA}}.

\vspace{-0.5cm}
\subsection{Experimental Results and Analysis}
\subsubsection{The performance of our methods}
\label{sec:ex_withtruth}
We evaluated our proposed multimodal semantic communication framework against two baseline scenarios: (1) classification using only video or audio, and (2) traditional communication methods. The traditional approach involved separate source and channel coding: converting video into frames with JPEG compression and using pulse-code modulation (PCM) for audio, with LDPC codes for channel coding.
\begin{table*}[!h]
    \centering
    \caption{Localization Accuracy Comparison across Different SNR, Channel Types, and with/without Trustworthy Encoder/Decoder (TC).}
    \label{tab:localization_accuracy_wRS}
    \begin{tabularx}{\textwidth}{|>{\raggedright\arraybackslash}X|>{\centering\arraybackslash}X|*{11}{>{\centering\arraybackslash}X|}}
        \hline
        \multirow{2}{*}{\textbf{Channel}} & \multirow{2}{*}{\textbf{TC}} & \multicolumn{11}{c|}{\textbf{SNR Values}} \\ 
        \cline{3-13}
        & & \textbf{0dB} & \textbf{3dB} & \textbf{6dB} & \textbf{9dB} & \textbf{12dB} & \textbf{15dB} & \textbf{18dB} & \textbf{21dB} & \textbf{24dB} & \textbf{27dB} & \textbf{30dB} \\ 
        \hline
        \multirow{2}{*}{AWGN} & w/o& 0.613 & 0.647 & 0.684 & 0.695 & 0.718 & 0.731 & 0.731 & 0.735 & 0.735 & 0.737& 0.737\
        \\\cline{2-13}
        & w & \textbf{0.643} & \textbf{0.660} & 0.684 & \textbf{0.708} & 0.718 & 0.731 & 0.731 & 0.735 & \textbf{0.736} & 0.737& 0.737\\
        \hline
        \multirow{2}{*}{Rayleigh} & w/o& 0.537 & 0.567 & 0.598 & 0.633 & 0.658 & 0.674 & \textbf{0.699} & 0.705 & 0.704 & 0.729 & 0.738 \\
        \cline{2-13}
        & w  & \textbf{0.560} & 0.567 & \textbf{0.619} & 0.633 & 0.658 & 0.674 & 0.683 & 0.705 & \textbf{0.706} & 0.729 & 0.738\\
        \hline
        \multirow{2}{*}{Rician} & w/o& 0.184 & 0.212& 0.25 & 0.430 & 0.433 & 0.611 & \textbf{0.714} & 0.717 & 0.721 & 0.731 & 0.733\\
        \cline{2-13}
        & w& \textbf{0.284} & \textbf{0.290} & \textbf{0.29} & \textbf{0.432} & 0.433 & 0.611 & 0.712 & 0.717 & \textbf{0.730} & 0.731 & 0.733 \\
        \hline
    \end{tabularx}
\end{table*}

\noindent \textbf{Performance Comparison}: Fig.~\ref{fig:accuracy} shows the relationship between localization accuracy and SNR across various channels—Additive White Gaussian Noise (AWGN), Rayleigh fading, and Rician fading. The results highlight the following:
\begin{itemize}
    \item Single-Modality Methods: These methods, which rely on either video or audio alone, exhibit lower accuracy due to missing modality information and lack of cross-modal guidance.
 \item Traditional Methods: Perform similarly to single-modality methods but with limited effectiveness in handling complex channel conditions.
 \item Proposed Multimodal Approach: Significantly outperforms single-modality and traditional methods. It achieves near-optimal localization accuracy under high SNR conditions and maintains robust performance even under low SNR and multipath fading scenarios. 
\end{itemize}
\noindent \textbf{Impact of Channel Conditions:} Our method demonstrates substantial resilience to noise interference, maintaining localization accuracy above 0.55 even at low SNR. As SNR increases, performance improves rapidly, showcasing the framework’s ability to handle multipath fading effectively. Overall, our multimodal framework exhibits enhanced effectiveness and robustness, offering superior performance in complex and high-noise communication environments compared to conventional techniques.

\subsubsection{The performance of our Multimodal Trustworthy Semantic Communication}
\label{sec:ex_withmodel}

We begin by illustrating an example of the multimodal task AVE in Fig.~\ref{fig:heatmap}. In this example, predicting the event is challenging due to the variability of the visual images and the presence of background noise in the audio signals. 

While both the stable AVE methods (AVEL~\cite{ave} and PSP~\cite{psp}) and our approach utilize AGVA, we demonstrate that our method provides improved attention to visual regions closely related to sound sources. As shown in Fig.~\ref{fig:heatmap}, For the event of a dog barking, our method focuses on the dog, particularly in the initial and final segments. In contrast, AVEL only identifies the background and has very limited receptive fields. Although PSP considers inter-modal enhancement, it overlooks the semantic and temporal variations of audio-visual data in the time-frequency domain. This synchronous mining should be effectively achieved through clear separation and lossless aggregation in the time-frequency domain. In the second row of the figure, it is evident that although PSP shows similar classification results to AVEL, it shares the same issue of focusing on the background and having limited receptive fields.
 
\subsubsection{The performance of our methods without RS encoding}
\label{sec:ex_with encoding}
To validate the effectiveness of MMTrustSC, we conducted an ablation experiment. This experiment assesses the model’s performance without the use of RS encoding and decoding. By maintaining the same semantic and channel encoding at the transmitter (i.e., consistent pre-trained models), we retrained the receiver’s model to handle multimodal tasks. As shown in Table ~\ref{tab:localization_accuracy_wRS}, we compared the experimental results of single-modal semantic transmission and multimodal semantic transmission from Fig.~\ref{fig:accuracy}. The accuracy of each model type under different SNRs clearly indicates a performance drop without RS encoding and decoding, especially at low SNR. This is because at low SNR, the impact of noise is greater, and RS encoding can correct a certain number of error bits, leading to a performance loss at low SNR when RS encoding is not used. This indicates that existing semantic communication methods have shortcomings in terms of trustworthiness, and errors generated after being subjected to attacks can affect the quality of communication, making multimodal trustworthy semantic communication critically important.

\section{Conclusion}
In this letter, we introduce MMTrustSC, a robust framework designed to enhance the security and reliability of multimodal semantic communication, particularly for Audio-Visual Event (AVE) localization tasks. By integrating advanced semantic encoding and decoding techniques, MMTrustSC effectively manages inter-modal complementarity and mitigates noise in complex environments, ensuring the integrity and privacy of transmitted data. Comprehensive semantic encryption at the transmitter and decryption at the receiver safeguard against unauthorized access, leading to significant improvements in transmission accuracy and reliability. Simulations demonstrate that MMTrustSC outperforms traditional methods, establishing a new standard for secure multimodal semantic communication.

\bibliographystyle{IEEEtran} 
\bibliography{our}

\vfill

\end{document}